\documentclass[aps,pra,twocolumn,superscriptaddress,a4paper]{revtex4}
\usepackage{amsmath}
\usepackage{graphicx}
\usepackage{ulem}
\usepackage{color}

\begin{document}

\newcommand{\abs}[1]{\left|#1\right|}
\newcommand{\ket}[1]{\left|#1\right>}
\newcommand{\bra}[1]{\left<#1\right|}

\title{Heralded generation of entanglement with coupled cavities}
\author{Jaeyoon Cho}
\affiliation{Department of Physics and Astronomy, University College London, Gower St., London WC1E 6BT, UK}
\author{Dimitris G. Angelakis}
\affiliation{ Centre for Quantum Technologies, National University
of Singapore, 2 Science Drive 3, Singapore 117542}
\affiliation{Science Department, Technical University of Crete,
Chania, Crete, Greece, 73100}
\author{Sougato Bose}
\affiliation{Department of Physics and Astronomy, University College London, Gower St., London WC1E 6BT, UK}

\date{\today}

\begin{abstract}
We propose a scheme to generate two-photon, two-atom, or atom-photon entangled states with a coupled system of two cavities. In our scheme, two cavity photons are exchanged by the direct inter-cavity coupling, while atoms in the cavities simply play the role of generating and probing them. By virtue of the high efficiency of atomic state measurement, this method enables the realization of efficient heralded entanglement generation robust against photon loss, which greatly facilitates applications in quantum information processing.
\end{abstract}


\maketitle

Entanglement is one of the essential ingredients of quantum information science. While various physical systems have exhibited entanglement, entangled photons have found unique applications in quantum communication between distant parties, such as quantum cryptography \cite{e91} and quantum teleportation \cite{bbc93}, thanks to their high portability.  Because of its relative ease of implementation, spontaneous parametric down-conversion has long been a conventional source of entangled photons \cite{s03}, and has allowed us to carry out various proof-of-principle experiments, such as the test of quantum nonlocality \cite{b64} and demonstration of two-qubit gates \cite{pittman03} and multipartite entanglement \cite{lu07}. The long-lived electronic ground states of single atoms, on the other hand, are favored for storing and manipulating local quantum information. There have been numerous proposals and experiments to implement conditional gates and generate entanglement of single atoms. To name a few, entanglement of single atoms were generated in ion traps by exploiting the collective vibrational mode as a quantum bus \cite{haffner05} and in optical lattices by exploiting collisional phase shifts \cite{mandel03}.

Another line of effort has been directed toward interfacing between single photons and single atoms to take advantage of both systems. A reasonable choice for doing that by is exploiting cavity QED to amplify otherwise weak coupling between them. For instance, cavity QED allows one to entangle a photon with an atom, and to map an atomic state into a photon \cite{raimond01}. Recently, these operations were combined to generate two-photon entangled states with an atom in a cavity \cite{wwk07}. One of the ultimate goals in this direction is to connect multiple cavities so that they can communicate with each other through intermediate photons. Such coupled-cavity systems have been considered as basic building blocks toward a scalable architecture for quantum information processing \cite{p97,asy07,amb07}, and recently also considered for quantum simulation \cite{hbp06}. The simplest case of coupled-cavity systems, i.e., that of two cavities each having a single atom, would be ideal for the generation of Bell-type entanglement, which is of great use in various modes of quantum information processing. However, most of the two-atom entanglement schemes suffer from loss of the intermediate photons leading to degradation of the final quality of entanglement. It would thus be desirable to have a heralded method for generating entanglement, where the cases of photon loss can be eliminated by heralding. This could be achieved by performing (non-deterministic) Bell-state measurement of two photons leaking out of cavities \cite{duan03,lim06}, or by performing polarization measurement of a photon reflected sequentially from cavities \cite{cho05}. These schemes, however, rely on single-photon measurement, which is still far from efficient. Moreover, a scheme for generating two-photon entanglement in such a system is still missing, except for a trivial extension of two-atom entanglement schemes, i.e., mapping entanglement of two atoms into two photons. In contrast to the case of a single cavity \cite{wwk07}, this system would emit entangled photons into different spatial modes, facilitating their use for other applications.

In this paper, we introduce a conceptually different mechanism for generating two-photon, two-atom, or atom-photon entangled states with a coupled system of two cavities each having a single atom. The brief idea is as follows. We first load two cavities each with a single photon having an orthogonal polarization. Both photons are then subjected to free inter-cavity hopping, which can be described by a beamsplitter-like transformation \cite{hbp06}. In a specific time, both photons will be distributed over two cavities, where the photons are in a superposition state of four possibilities, two of which are with both photons occupying only one cavity and the other two with one photon per cavity. At this instant, each cavity mode is probed by the atom in such a way that the resulting atomic state indicates whether the cavity mode was empty or not. Only when the ensuing measurement of the atomic state reveals that neither of the cavities is empty, we take the state, whereby the polarization-entanglement is diverted from the superposion state of the cavities. According to the way of probing, we end up with two-photon, two-atom, or atom-photon entanglement.

A remarkable property of this approach is that the success of entanglement generation is heralded by measurement of atoms, which is known to be efficient. Moreover, the quality of the generated entanglement heavily relies on that of the measurement and any dissipation prior to the measurement can be detected. The heralded entangled state is thus guaranteed to be of high quality. This kind of entanglement source is actually preferable, especially in quantum cryptography \cite{grt02} and linear optics quantum computation \cite{dhn06}. Moreover, a redundant array of heralded entanglement sources accompanied with appropriate feedforward will asymptotically serve as a deterministic source of entangled states.

\begin{figure}
\includegraphics[width=0.64\columnwidth]{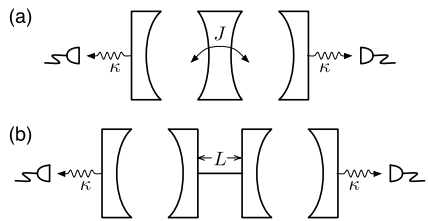}
\caption{Coupled systems of two cavities. Two cavity modes are coupled either (a) directly or (b) via a fiber. $J$ and $\kappa$ denote the inter-cavity hopping rate of photons and decay rate of photons into the output channel, respectively. By tailoring both the fiber length $L$ and the resonant frequency of the cavity, system (b) reduces to (a).}
\label{fig:setup}
\end{figure}

The system at hand consists of two single-mode cavities having the same resonant frequency, which are coupled to each other. There are two considerable cases. FIG.~\ref{fig:setup}(a) shows the first case in which the coupling is achieved by the overlap of evanescent fields out of the intermediate cavity mirror. In this case, the interaction Hamiltonian for cavity photons is simply given by (without cavity decay)
\begin{equation}
H_I=J(a^\dagger b + b^\dagger a),
\label{eq:hamiltonian}
\end{equation}
where $a$ and $b$ are the annihilation operators for the two cavity modes, respectively, and $J$ is the rate of inter-cavity hopping of photons \cite{hbp06}. The other case is where the coupling is mediated by a quantum channel such as a fiber (or simply vacuum), as shown in FIG.~\ref{fig:setup}(b). In this case, the interaction Hamiltonian in the rotating frame can be written as
$H_F=\sum_n \nu_n[a^\dagger f_n + (-1)^n b^\dagger f_n + h.c.] + \sum_n \Delta_n f_n^\dagger f_n$,
where $f_n$ is the annihilation operator for the $n$th fiber mode, $\nu_n$ is the coupling rate of the cavity mode to the $n$th fiber mode, and $\Delta_n$ is the frequency difference between the $n$th fiber mode and the cavity mode \cite{p97}. Here, the factor $(-1)^n$ accounts for the phase difference between adjacent modes at the fiber end. We confine our interest to the case of a short fiber, wherein the fiber mode is highly discrete. Moreover, we assume that no fiber mode is resonant to the cavity mode and the nearest mode is far-detuned, that is, the minimum of $\abs{\Delta_n}$ is much larger than the coupling rate $\nu_n$. In this regime, excitation to the fiber mode is highly suppressed and the Hamiltonian reduces to Eq.~(\ref{eq:hamiltonian}) by adiabatic elimination, with an effective inter-cavity coupling rate given by $J=\sum_n(-1)^n\nu_n^2/\Delta_n$. Note that this regime is within the reach of current technology. For example, if we take $L=1~\text{cm}$ and $\kappa_c/2\pi=10~\text{MHz}$, where $L$ is the length of the fiber and $\kappa_c$ is the decay rate of the cavity into continuum modes, the mode spacing of the fiber $(\Delta_{n+1}-\Delta_n)/2\pi=c/2L=15~\text{GHz}$, where $c$ is the speed of light, is found to be much larger than the cavity-fiber mode coupling rate $\nu_n/2\pi\sim\sqrt{(\kappa_c/2\pi)(c/L)}=0.55~\text{GHz}$. In what follows, we shall consider Eq.~(\ref{eq:hamiltonian}) as our model Hamiltonian.

\begin{figure}
\includegraphics[width=0.36\columnwidth]{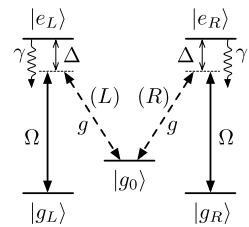}
\caption{Involved atomic levels and transitions. Two transitions are coupled, respectively, to two orthogonally polarized modes of the cavity with coupling rate $g$ and detuning $\Delta$. The amount of $\Delta$ is adjusted by an additional laser. Ground levels $\ket{g_{L}}$ and $\ket{g_{R}}$ are decoupled from the cavity mode. $\Omega$ denotes the Rabi frequency of the classical field, which is needed only for the generation of atomic entanglement, and $\gamma$ denotes the atomic spontaneous decay rate. Additional classical lasers required for preparation and measurement of the atomic state are omitted in this figure.}
\label{fig:atom}
\end{figure}

We first consider the generation of two-photon polarization entangled states.
The first step toward the entanglement generation is to load two cavities, respectively, with orthogonally polarized single photons $\ket{L}$ (left-circular polarization) and $\ket{R}$ (right-circular polarization). For this, two single atoms are introduced, one into each cavity. The present scheme can be applied to both cases of trapped atoms and flying atoms \cite{yvk99,raimond01,wsb04}. For convenience, we explain the scheme assuming the atoms are trapped in the cavities, which is more straightforward to understand. FIG.~\ref{fig:atom} depicts the atomic level structure we consider. The transition between ground state $\ket{g_{0}}$ and excited state $\ket{e_L}$ ($\ket{e_R}$) is coupled to the left-circularly (right-circularly) polarized mode of the cavity with coupling rate $g$ and detuning $\Delta$. Other ground states $\ket{g_{L}}$ and $\ket{g_{R}}$ are decoupled from the cavity mode. We assume that although the atom is initially far-detuned ($\Delta\gg g$), the detuning can be controlled by ac Stark shift induced by strong classical fields. In the case of generation of entangled photons, the Rabi frequency $\Omega$ of the classical field is not taken into account. In order to generate single photons $\ket{L}$ and $\ket{R}$ in the cavities, two atoms are initialized, respectively, into excited states $\ket{e_L}$ and $\ket{e_R}$, and the detuning $\Delta$ is adjusted to zero for a period of time $\pi/2g$, allowing resonant interaction between the atom and the cavity. For this process not to be disturbed by other transition channels of the cavity photon, we require the strong atom-cavity coupling regime $g\gg J, \kappa, \gamma$, where $\kappa$ is the decay rate of the cavity photon into the output channel (see FIG.~\ref{fig:setup}) and $\gamma$ is the spontaneous decay rate of the atom.

Once the cavity photons are prepared, the second step is to turn off the classical fields applied for the ac Stark shift in the first step, so that the detuning $\Delta$ is returned to the initial value. Since the atomic transition is far-detuned, the Hamiltonian conditional on no cavity decay is now expressed by the summation of the inter-cavity hopping terms [Eq.~(\ref{eq:hamiltonian})] and the cavity decay terms:
\begin{equation}
H=J\sum_{p=L,R}(a_p^\dagger b_p+b_p^\dagger a_p)-i\frac{\kappa}{2}\sum_{p=L,R}(a_p^\dagger a_p+b_p^\dagger b_p),
\end{equation}
where the subscripts represent the polarization. By changing the basis as $x_p=\frac{1}{\sqrt2}(a_p+b_p)$ and $y_p=\frac{1}{\sqrt2}(a_p-b_p)$, this Hamiltonian takes a simple form:
\begin{equation}
H=J\sum_{p=L,R}(x_p^\dagger x_p-y_p^\dagger y_p)-i\frac{\kappa}{2}\sum_{p=L,R}(x_p^\dagger x_p+y_p^\dagger y_p),
\label{eq:hamil2}
\end{equation}
and the photonic state $a_L^\dagger b_R^\dagger \ket0$ is written as
\begin{equation}
\ket{\Psi(0)}=\frac{1}{\sqrt2}(\ket{\Phi^-}+\ket{\Psi^-}),
\end{equation}
where $\ket{\Phi^-}=\frac{1}{\sqrt2}(x_L^\dagger x_R^\dagger - y_L^\dagger y_R^\dagger)\ket0=\frac{1}{\sqrt2}(a_L^\dagger b_R^\dagger+a_R^\dagger b_L^\dagger)\ket0$ and $\ket{\Psi^-}=-\frac{1}{\sqrt2}(x_L^\dagger y_R^\dagger - x_R^\dagger y_L^\dagger)\ket0=\frac{1}{\sqrt2}(a_L^\dagger b_R^\dagger-a_R^\dagger b_L^\dagger)\ket0$.
Note that the state $\ket{\Psi^-}$, which is a two-photon polarization entangled state, is invariant under the inter-cavity hopping part in Hamiltonian~(\ref{eq:hamil2}). Consequently, once this state is prepared in the cavity, the output photons are guaranteed to remain in the same entangled state with a definite pulse shape. To this end, we are interested in the (unnormalized) conditional state at time $\pi/4J$:
\begin{equation}
\ket{\Psi\left(\frac{\pi}{4J}\right)}_C=\frac{e^{-\frac{\pi\kappa}{4J}}}{\sqrt2}(-i\ket{\Phi^+}+\ket{\Psi^-}),
\label{eq:state}
\end{equation}
where $\ket{\Phi^+}=\frac{1}{\sqrt2}(x_L^\dagger x_R^\dagger + y_L^\dagger y_R^\dagger)\ket0=\frac{1}{\sqrt2}(a_L^\dagger a_R^\dagger+b_L^\dagger b_R^\dagger)\ket0$.
Note that the state $\ket{\Phi^+}$ is such that only one cavity has both the photons while the other cavity is empty. This is clearly distinguished from the state $\ket{\Psi^-}$ characterized by one photon per cavity.  Consequently, if we perform a non-demolition measurement distinguishing between zero and one photon at each cavity and take the state only when both cavities have one photon, we can extract the state $\ket{\Psi^-}$. If the measurement is ideal, it succeeds with probability $P=\frac{1}{2}\exp\left(-\frac{\pi\kappa}{2J}\right)$, which approaches $\frac12$ as $J/\kappa$ increases. Remarkably, regardless of the success probability, the resulting state has in principle unit fidelity.

The remaining question is how to perform the non-demolition measurement distinguishing between zero and one photon. We also require the measurement to be achieved without distinguishing between polarizations and to be fast enough for the photonic state~(\ref{eq:state}) not to evolve during the measurement. Such a measurement is again aided by the atoms inside the cavities \cite{raimond01,devitt07}. Before starting the measurement, each atom should be prepared in state $\ket+\equiv\frac{1}{\sqrt2}(\ket{g_L}+\ket{g_0})$. Recalling that the atoms are in state $\ket{g_{0}}$ as a result of the single-photon generation in the first step, this can be easily done by classical Raman pulses. Note that the preparation can be performed without disturbing the evolution of photons during the second step, since the atoms are far-detuned from the cavity mode. The detuning $\Delta$ is now adjusted to zero, allowing resonant atom-cavity interaction, for a period of time $\pi/g$. From FIG.~\ref{fig:atom}, it is easily seen that if there were one photon in a cavity, the resulting state would be  $\ket-\equiv\frac{1}{\sqrt2}(\ket{g_L}-\ket{g_0})$ with a remaining single photon having the same polarization. If there were no photon, however, the atomic state would not be changed. We can thus distinguish between the two cases by measuring the resulting atoms in the $\ket\pm$ basis. In the case of two photons, the resulting atomic state contains excited-state components as well, thus the atom could be measured in both states. This does not arise as a problem, however, since in that case the other cavity should be empty. To sum up, the generation of two-photon entanglement succeeds only when both atoms are measured in state $\ket-$. Note that this measurement is fast because we are assuming the strong atom-cavity coupling regime.

The two-atom entangled state can also be generated in a similar manner just by modifying the above non-demolition measurement step as follows. In this case, we start from the state $\ket{g_{0}}$, which is automatically prepared by the single-photon generation step. Right after the second step, the detuning $\Delta$ is adjusted to zero for a period of time $\pi/\sqrt{2}g$, during which the classical field with Rabi frequency $\Omega$ in FIG.~\ref{fig:atom} is also applied with $\Omega=g$. It is easily seen that if the cavity was in the single-photon state $\ket{L}$ ($\ket{R}$), this operation coherently transfers the atomic state completely into state $\ket{g_{L}}$ ($\ket{g_{R}}$). On the other hand, if the cavity was empty, this operation does not change the atomic state. Consequently, if the resulting atomic state is measured by observing the resonance fluorescence on a transition between $\ket{g_{0}}$ and an excited state, the zero-photon state can be distinguished from the single-photon state, while at the same time the photonic state is mapped into the atom. By discarding the state with an empty cavity, the atoms thus remain in an entangled state $\frac{1}{\sqrt{2}}(\ket{g_{L}}\ket{g_{R}}-\ket{g_{R}}\ket{g_{L}})$. Note that by replacing the state mapping with the previous non-demolition measurement in either of the cavities, this scheme is straightforwardly extended to the case of atom-photon entanglement generation. A further extension would be combining entangled atom pairs to generate a multi-qubit cluster state \cite{lim06}.

In the remainder of this paper, we discuss the effects of decays on the performance of the scheme. Since the Hilbert space of the system at hand is not small enough to be dealt with by exact analytic calculations, we take the perturbation approach and obtain the involved states up to first order in small constants $J/g$, $\kappa/g$, and $\gamma/g$. Let us first consider the generation of cavity photon $\ket{L}$, which is achieved by having an atom in state $\ket{e_L}$ interact with the resonant cavity mode for a period of time $\pi/2g$. This process is disturbed by the atomic spontaneous decay, the cavity decay, and hopping of the photon into the other cavity. Among them, the first two, which result in photon loss, do not affect the fidelity of the final entangled state, since loss of photons can be detected by the measurement. Conditional on having one photon, the final state of this process is found to be $(a_L^\dagger-i\frac{J}{g}b_L^\dagger)\ket{0}\ket{g_1}$ up to first order. Here, we neglected the effect occurring when the photon having hopped into the other cavity is absorbed by the other atom, since it is a higher-order contribution. This state is approximately the same as the state we get when the second step of the scheme proceeds for a time $\tau$ given by $J\tau=J/g$. This amount of time can be thus compensated by decreasing the interaction time for the second step. Assuming the pulse timing is exact up to first order, we can regard the state of the cavity photons just before the final step as being nearly perfect. Let us first consider the case of the two-photon entanglement generation. The non-demolition measurement succeeds only when both atoms prepared in state $\ket+$ are measured in state $\ket-$. Up to first order, this process completely filters out the case of having two photons in one cavity and no photon in the other cavity, since the probability that the photon hops to the empty cavity and flips the atomic state is of higher order. In case each cavity has one photon, however, the photon loss could lead to an erroneous measurement result due to the loss of atomic coherence. If both atoms are measured in state $\ket-$, the final state of the photons $\rho_P$ is proportional to  $(1-2P_{L})\ket{\Psi^-}\bra{\Psi^-}+P_L\cdot\frac14 I\otimes I$, where $P_{L}=\frac{\pi}{4g}(\gamma+\kappa)$ and $I$ is the identity operator. Here, for simplicity, we have assumed that when photons are lost, the state is fully mixed. The fidelity of the final state $\bra{\Psi^-}\rho_P\ket{\Psi^-}$ is thus given by
$F_{P}=1-(3\pi/16)\cdot(\gamma/g+\kappa/g)$.
Note that inaccurate control of the detuning, i.e., a nonzero $\Delta$, rather increases the fidelity by suppressing the spontaneous decay, although it decreases the success probability.
In the case of two-atom entanglement generation, cavity decay during the final step can be detected, since it leaves the atom in state $\ket{g_{0}}$. Some portion of the atomic spontaneous decay, i.e., decay into state $\ket{g_{0}}$, is also detected, according to its branching ratio. For simplicity, we assume the branching ratio of the decay to $\ket{g_{0}}$ and $\ket{g_{L,R}}$ is 50:50. A similar calculation yields the fidelity of the two-atom entangled state as
$F_{A}=1-(3\pi/16\sqrt2)\cdot(\gamma/g)$.
Recalling that we are assuming the strong atom-cavity coupling regime $g\gg\gamma,\kappa$, these fidelities are reasonably high. There are several cavity models which are expected to exhibit very strong coupling regimes \cite{skv05}. Even in a moderately strong coupling regime $g/10\sim\gamma,\kappa$ \cite{yvk99}, if we take the state only when both cavities output a photon, as in most current experiments based on single photons, the fidelity of the two-photon entangled state becomes $F_{P}=1$ up to first order.

This work has been supported by the Korea Research Foundation Grant (KRF-2007-357-C00016) funded by the Korean Government (MOEHRD), QIP IRC (GR/S821176/01),  and the
European Union through the Integrated Projects SCALA (CT-015714).
 SB would like to thank the Engineering and Physical Sciences Research
Council (EPSRC) UK for an Advanced Research Fellowship the support
of the Royal Society and the Wolfson foundation.

\end{document}